# Single-electron latch with granular film charge leakage suppressor

Alexei O. Orlov, Xiangning Luo, Kameshwar K. Yadavalli[*], Igor S. Beloborodov[†],

and Gregory L. Snider

Dept. of Electrical Engineering, University of Notre Dame, Notre Dame, IN 46556

**Abstract-** A single-electron latch is a device that can be used as a building block for Quantum-dot Cellular Automata (QCA) circuits. It consists of three nanoscale metal "dots" connected in series by tunnel junctions; charging of the dots is controlled by three electrostatic gates. One very important feature of a single-electron latch is its ability to store ("latch") information represented by the location of a single electron within the three dots. To obtain latching, the undesired leakage of charge during the retention time must be suppressed. Previously, to achieve this goal, multiple tunnel junctions were used to connect the three dots. However, this method of charge leakage suppression requires an additional compensation of the background charges affecting each parasitic dot in the array of junctions. We report a single-electron latch where a granular metal film is used to fabricate the middle dot in the latch which concurrently acts as a charge leakage suppressor. This latch has no parasitic dots, therefore the background charge compensation procedure is greatly simplified. We discuss the origins of charge leakage suppression and possible applications of granular metal dots for various single–electron circuits.

[*] Dept. of Electrical Engineering, 56-125B Engineering IV Building, University of California Los Angeles, Los Angeles, CA 90095, USA

[†] Materials Science Division, Argonne National Laboratory, Argonne, IL 60439, USA; James Franck Institute, University of Chicago, IL 60637, USA

**Introduction.**

Quantum-dot Cellular Automata (QCA)[1,2] is an inherently nanostructure-compatible binary architecture that uses interaction (electrostatic or magnetic) between nominally identical building blocks ("cells") to encode and process binary information with minimal power dissipation. By modulating the potential barriers using clocking field, power gain, pipelining, and very low power dissipation can be achieved in QCA [3]. Both electronic and magnetic versions of clocked QCA were recently experimentally demonstrated [4,5]. An electronic version of clocked QCA uses "single-electron latches" (or "QCA half-cells") as building blocks to assemble larger QCA arrays.

A single-electron latch (sometimes also referred to as "parametron"[6]) is an elementary single-electron logic device where a binary bit of information is represented by the spatial location of an electron within the device. It has three input terminals (differential signal inputs $V_{IN}^+$ and $V_{IN}^-$, and clock input $V_{CLK}$) and two differential output terminals $V_{OUT}^+$ and $V_{OUT}^-$. There are three possible states of the latch: (1) "null", when it stores no information (and is electrically neutral); (2) "active," when switching of an electron takes place; and (3) "hold," when binary information is preserved in the form of differential output voltage (positive or negative for "0" or "1") which then acts as an input for the next latch in the circuit.

In its metal-tunnel junction implementation[6,7] a latch consists of three metal "dots" separated by tunnel junctions (Fig. 1(a)), where the tunneling through the center dot D2 can be controlled by the clock bias $V_{CLK}$, so that D2 acts as an adjustable Coulomb barrier separating the end dots. This scheme of single-electron control can also be used for molecular QCAs, where metal dots will be replaced with molecular complexes [8].

The operation sequence of a single-electron latch is defined by the combination of applied clock and input voltages. Initially, both clock and input signals are set to zero, so the latch is electrically neutral (null state). In this state it does not carry any information. When a small differential input voltage, $V_{IN}$, is applied to end dots (D1 and D3) the latch remains in the null state because the input signal is insufficient to lift the Coulomb blockade for tunneling within the dots. Only when the clock signal, $V_{CLK}$, is applied is an electron forced from D2 to the end dot coupled to positive input, $+V_{IN}$. The magnitude of $V_{CLK}$ is chosen to maximize the Coulomb blockade for tunneling from the end dot after a single electron is transferred, thus trapping ("latching") an electron. Once the electron is latched, the input signal can be removed, and the electron remains trapped until $V_{CLK}$ is set back to zero.

However, once an electron is latched and input signal is removed, there is a finite probability for it to escape to the opposite end dot (it cannot return to the middle dot since it is energetically unfavorable). The escape of an electron during the hold time of the latch is called a "leakage" or "decay" error, which is characterized by leakage error rate, $\Gamma_L$ ($s^{-1}$). There are several mechanisms which can cause this process: (1) thermal excitation over the Coulomb barrier; (2) simultaneous tunneling of several electrons (cotunneling[9]); (3) photon-assisted tunneling and cotunneling (PAT)[10]; and (4) random background charge fluctuations (RBC).

One way to reduce the leakage rate in single-electron devices is to increase the number of junctions N connecting the dots. In single-electron devices such as pumps[11] and traps[12] the use of multiple tunnel junctions (MTJs) reduces the leakage rate because each of the above leakage mechanisms is a strong function of the number of junctions N.

An increase in N raises the height (W) of the Coulomb barrier[13] (W ∝ N) thus reducing the leakage rate due to the thermal activation : $\Gamma_{LTherm} \sim \exp(-W/k_B T)$. It also reduces the leakage rate due to cotunneling[9] $\Gamma_{LCot} \propto \left[\frac{(eV)^2 + (k_B T)^2}{E_C^2}\right]^{N-1}$ under the influence of small bias V« $E_C$/e across the array of N junctions ($E_C$ is charging energy) for non-zero temperature, *T*. The rate of photon-assisted electron leakage[14], $\Gamma_{LPAT}$, is also reduced because it is inversely proportional to the barrier height: ($\Gamma_{LPAT} \propto 1/W$). Leakage processes in single-electron devices containing MTJs were extensively studied in recent publications[14-21]. Experimental and theoretical investigations[20, 22] suggest that in metal single-electron devices with MTJs at moderately high temperatures (≥150 mK) or in the presence of a bias across the junctions (V≥50 μV) the leakage errors are dominated by thermal activation over the Coulomb barrier [20, 23-25]. At lower temperatures (<100 mK) and for N<5 cotunneling is deemed to be the dominant mechanism of leakage errors [26], and for N ≥5 a large number of experiments suggest that leakage errors are dominated by photon-assisted tunneling and cotunneling driven by non-equilibrium *1/f* noise associated with heat dissipation in the substrate[27]. Theory and experiments show that this noise has enough power at microwave frequencies to trigger the PAT processes[14, 19]. These results explain the large discrepancy (by many orders of magnitude) between the observed leakage rate and that predicted by the orthodox theory of single-electron tunneling and cotunneling[9, 13] for metal tunnel junction devices. Finally, leakage errors could be caused by the RBC fluctuations which affect the long-term stability of the single-electron devices. The RBC fluctuations change the potential profile around the device and therefore require re-tuning of the entire device (since they affect the entire device operation and not just the leakage errors). These fluctuations occur on the timescale of the order of hours. The RBC fluctuations and photon-assisted tunneling apparently have the same physical origin (relaxation of non-equilibrium traps resulting in non-equilibrium 1/f noise) and tend to decay in time if the device is kept at cryogenic temperatures for long periods of time (»100 hours) [27, 28].

To reduce the leakage errors in a latch at low temperatures (<100 mK) and without a bias between the end dots (hold state) occurring within the stability time when the system remains immune to RBC fluctuations (~1000 seconds in our experiments), the dominant error mechanism, i.e. photon-assisted tunneling, must be suppressed. For the four junction (N=4) traps[12] the experimentally determined leakage rate was found to be $\Gamma_L > 1$ s$^{-1}$; and in pumps with no gate bias applied the measured rates are $\Gamma_L \sim 10$ s$^{-1}$ (N=4) [14,28] and $\Gamma_L \sim 10^{-2}$ s$^{-1}$ (N=6)[14]. Experimentally determined leakage rate in the latch with N=6 [29] in the hold state was found to be $\Gamma_L \sim 2$ s$^{-1}$. The higher leakage rate in the latter experiment compared to six junction pump[14] is expected because the electrostatics for the latch in hold state is not the same as it is for a pump: the equivalent of hold state for the pump is a null state of the latch[7], when it carries no information; this is the ground state and is therefore more stable. The hold state in the latch is more similar to that of a single-electron trap in an equilibrium, where probabilities for trapping and escape are equal[23]. The major difference between the trap and a latch is how they store the trapped electron: while a trap is usually designed to store the charge after the gate bias is removed, a latch stores an electron while the clock signal is applied and must return to the initial neutral state once the clock signal is set to zero.

The fabrication of latches with MTJs has its own drawbacks, because it leads to the unavoidable formation of additional "dots" that are affected by the random background charges. This requires additional compensation of these charges for each extra dot which drastically complicates the tuning and operation of the device.

One alternative way to suppress the leakage caused by cotunneling is to use resistive microstrips in series with tunnel junctions instead of increasing N[30]. The cotunneling current, $I_C$, in N-junction arrays in series with 2 resistors of value R in the Coulomb blockade regime obeys a power law [31],

$$I_C \propto V^\eta, \quad \eta = 2(N+z)-1 \qquad (1)$$

where V is the bias across the array; $R_Q = h/e^2$ is the resistance quantum, and $z = R/R_Q$ is a dimensionless parameter. The simple implication of (1) is that resistive strips with resistance $R = R_Q \times z$ act as MTJs with z junctions, thus reducing the cotunneling rate[32]. The use of embedded resistors reduces the number of junctions required which simplifies the problem of random background charge compensation for the additional dots of MTJs as well as the design of devices. Experiments have demonstrated the feasibility of the proposed design for electron traps (N=4) [33] and pumps (N=3) [26] where cotunneling sets the limits to the accuracy of charge transfer. It must be noted that it is extremely difficult to fabricate metallic resistive microstrips of small size, consequently increasing the minimum possible size of the device. Moreover, the use of such microstrips for latches is non practical due to large self-capacitance of the mictrostrips, ≈60 aF/μm[32] which would severely degrade the charging energy of the dots if such microstrips are embedded in the latch. The increase of the resistance per square can be obtained by using granular metals, however the theory[30, 31] is derived only for diffusive conductors where the charge equilibrium is established before a tunneling event occurs. The operation of a single electron transistor (SET) with granular metal microstrips connecting an Al island to the environment was studied in[34]. The observed strong non-linearity in the $I_{ds}(V_{ds})$ characteristics of the SET in the blockade region $I \sim V^{\alpha}$, $\alpha \approx 10$ suggests that the cotunneling was strongly suppressed. In our recent work[35], we demonstrated the operation of a single-electron transistor where a granular metal film was used as an island material and two traditional $AlO_x$ tunnel junctions connected the island to the source and drain wires. We showed that such a device also exhibits strong nonlinearity in the blockade region: $I \sim V^{\alpha}$ where $\alpha \approx 9$. It must be noted that the direct application of formula (1) would lead to a different conclusion based on estimate of parameter $\eta \gg 10$, which is not surprising since the theory was developed for linear diffusive resistors, whereas the conduction mechanism in granular film is attributed to hopping of electrons between the remote grains [36]. We speculate that this method of cotunneling suppression must also reduce the photon-assisted tunneling through the devices and therefore can be used in the design of single-electron latches. A strong nonlinearity of the I-V characteristic of the film is expected to increase the barrier for tunneling in the hold state. Also, the size of the granular metal dot can be made much smaller than that of diffusive linear resistors, so the self capacitance of such a dot can be made very small. To verify the relevance of this approach we fabricated and characterized a single electron latch with granular metal (GM) middle-dot connected to the two Al end dots by single $AlO_x$ junctions.

**Fabrication and measurement technique**.

The device was fabricated by two steps of e-beam lithography and shadow evaporation[37]. In the first e-beam lithography step the latch is fabricated, and in the second step two SET electrometers[4] are fabricated. The sketch of the device is shown in Fig. 1a. To fabricate the latch two Al end dots (D1 and D3 in Fig. 1a) and a "dummy" middle dot are first deposited at an angle of +6.2 degrees to the normal followed by an in-situ oxidation of the surface of the Al layer. Then, a $CrO_x$ GM film (the middle dot D2) is formed by evaporating Cr at a negative angle, -6.2 degrees to the normal in oxygen ambient. As a result, in the overlap region, we obtain a thin $AlO_x$ layer sandwiched between Al and $CrO_x$ layers. The resistivity of the GM strip is controlled by adjusting the oxygen pressure in the evaporator chamber [34, 38]. The sheet resistance of the GM strip is estimated from the measurements of an SET with CrOx island (Fig. 1b) fabricated along with the latch with the same design of the island and the junctions [35]; sheet resistance ~3-5 k$\Omega$/□ at room temperature is obtained. At low temperatures (~100 mK) the value of the sheet resistance of the GM strip with zero bias across it is on the order of 10 M$\Omega$/□, but it drops exponentially down to about 100 k$\Omega$/□ for 0.5 mV of applied bias. The value of the junction resistance was obtained from the measurements of the SET at high source–drain bias when the Coulomb blockade is completely suppressed ($eV_{ds} \geq 10\ E_C$), $R_J \sim 10$ M$\Omega$. Two Al-$AlO_x$ electrometers (E1 and E2), and the input and clock wires (Fig. 1a) are formed in the second layer e-beam lithography step followed by shadow evaporation of Al. Differential input signals (+$V_{IN}$ and -$V_{IN}$) and the clock signal, $V_{CLK}$, are capacitively coupled to D1, D3, and D2 respectively. The alignment tolerance between the two steps of e-beam lithography is about 300 nm according to the pattern design and an alignment accuracy of 200 nm between features defined in the two steps of e-beam lithography is achieved by using pre-defined alignment marks. For comparison, an Al/$AlO_x$ latch with the same design as above, except that the $CrO_x$ island is replaced by an Al island, is fabricated on the same substrate. In addition, an Al/AlO$_x$ latch with 4 junctions and similar size dots and coupling capacitors is also fabricated.

The low temperature transport measurements are performed in a dilution refrigerator with a base temperature of 15 mK. Standard lock-in techniques are used in all of the low temperature measurements with small excitation voltages (10-50 µV) applied to the electrometers to avoid excessive heating of the latch [18]. SET conductance is measured at a frequency ~3 kHz with time constant ~3 ms to provide adequate temporal resolution for phase detection of the pulses with duration ≥100 ms. The typical electron temperature of the device is estimated to be about 70 mK [4]. A magnetic field of 1 T was applied to suppress the superconductivity of Al.

**Experimental results**.

The major goal of the experiments described below is to determine the ability of the latch with GM middle dot to suppress charge leakage. The observation of bistability is a direct evidence of single electron latching that can only be observed when undesired tunneling is strongly suppressed[4], whereas a lack of bistability clearly points to an excessive leakage rate[39]. One way to observe bistable behavior is to measure "phase plots" in $V_{IN}$, $V_{CLK}$ coordinates for the two directions of input bias voltage[4, 39]. A phase plot is a 3-D plot obtained by measuring the voltage on D1 (D3) (Z axis) using the SET electrometer E1 (E2) while linearly scanning input (X axis) and stepping clock (Y axis) voltages. The electrometer is biased in the middle of the rising slope of a Coulomb blockade oscillation so that a positive voltage increment on D1 leads to an increase of conductance and vice versa. Black color on the plot corresponds to a negative voltage increment and white color corresponds to a positive voltage increment on the dot. Figures 2(a) and 2(b) show two phase plots acquired for different directions of the input bias scans for a latch with 2 junctions and $CrO_x$ middle dot. Bistability in the latch is clearly visible and marked by opposing triangular areas on the phase plot in Fig. 2(a) and Fig. 2(b). For comparison, a phase plot for an Al/AlO$_x$ latch with the same design (two junctions) and Al middle dot is shown in Fig. 2c (since the phase plots for two directions of the input voltage scans are indistinguishable, only one plot is shown). In this case we observe no signs of bistability. The transition from one charge state to the other (at $V_{IN} \approx$ 2 mV) appears smooth meaning that electron moves back and forth in the transitional region and electrometer measures the time average of the dot potential. The Al/AlO$_x$ latch with 4 junctions shows only hints of bistability indicated by abrupt transitions, but the bistable area was undeveloped and no latching was observed [40]. The results of these experiments indicate that in the latch with GM middle dot charge leakage is suppressed much strongly than in 2-junction and 4 junction latches having Al middle dots.

The functional operation of a latch with GM middle dot is demonstrated with the input and clock pulse sequence shown in Fig. 3 (a, b) (the detailed description of the setup for this test is given elsewhere [4, 39]). The electrostatic potential of dot D1, monitored by electrometer E1, is shown in Fig. 3(c). Here we concentrate on the leakage errors occurring when an electron is trapped on one of the end dots (by the application of the clock signal) and the differential input signal is removed (from $t_0$ to $t_1$ and from $t_2$ to $t_3$ in Fig. 3c). An instance of such an error is shown by a dashed line in Fig. 3c. To determine the leakage rate we repeated the latching sequence in Fig. 3(a, b) multiple times and then analyzed the statistical distribution of the cumulative number of leakage errors as a function of time (Fig. 3d). The probability of the leakage error changes in time as [29]:

$$P_{err}(t) = n_{err}(t)/n_{total} = 1 - \exp(-t/\tau) \qquad (2),$$

where $\tau$ is the retention time constant, $n_{err}(t)$ is the total number of leakage errors that happened within time interval t and $n_{total}$ is the total number of clock cycles. By fitting our data to (2), we extracted the value of $\tau = 1/\Gamma_L \approx 1s$. Hence the average electron leakage rate is $\Gamma_L \approx 1\ s^{-1}$. Thus we conclude that the presence of GM film reduces leakage error rate to the level observed in an Al/AlO$_x$ latch with 6 junctions[4].

Another remarkable feature of the latch with GM middle dot is the uniformity of the observed charging diagram pattern for large scale scans of clocking and input biases depicted in Fig. 4a. For comparison, the data obtained from the 6-junction Al/AlO$_x$ device[41] is shown in Fig. 4b. Due to the presence of the random background charges on the parasitic islands in the 6-junction Al/AlO$_x$ device, the pattern in Fig. 4b is non-uniform and only small areas in the plot (where the distinct triangular pattern is present) can be used for latch operation. In these areas the random background charges on the parasitic dots are compensated by particular combination of the offset voltages on the input and clock electrodes. A change in the RBC configuration beyond the stability time changes *the pattern* observed in Fig. 4b so that latching may not be achievable for the same areas where the triangular pattern is currently seen in the plot, requiring retuning of the latch operating conditions ($V_{IN}$, $V_{CLK}$). In contrast, *the pattern* in Fig. 4a experience no such change: the latch with GM middle dot shows uniform periodic bistable behavior where the periods are defined by the coupling capacitors; the change in the RBC configuration leads only to an offset of the pattern as a whole along axes $V_{CLK}$ and $V_{IN}$. Therefore, it can be easily compensated by small adjustments of the offset biases applied to the input and clock electrodes, as there are no extra parasitic dots which require individual compensation. This result correlates with the observation that the gate voltage dependence of the conductance for an SET with GM island[35] only exhibits single period of Coulomb blockade oscillations, i.e. the island behaves as a uniform metal dot.

**Discussion.**

Two pertinent questions will be addressed here: (1) why is it possible to achieve control over single electron charge transfer in a latch with GM middle dot, and (2) what causes the suppression of charge leakage in such a device?

The GM film forming the middle dot in the latch can be viewed as an array of interconnected nanometer-sized metal dots intermixed with the oxide. The unavoidable presence of the background charge in the film randomly shifts the Coulomb blockade thresholds for the individual grains affecting their charging energy such that there is no hard gap in the density of states in the GM dot[42]. The value of capacitance coupling the gate to an individual grain can be estimated based on the average grain size ~10 nm[35], and it is orders of magnitude less than the capacitance from the gate to the whole GM dot, so that the periods of Coulomb blockade oscillations are expected to be far greater for the individual grains than for the whole GM dot. As a result, a change in the gate voltage does not influence individual grains ( for the same reason frustrated 2D arrays of tunnel junctions studied in[43] exhibit no gate voltage dependence) and the GM dot behaves as a metal with respect to the external gate despite the fact that its conductance exhibits non-metallic temperature dependence[35]. This makes it possible to control the electron transfer between the end dots in the single-electron latch with $V_{CLK}$ applied to the GM middle dot.

As regards the suppression of charge leakage, the presence of an MTJ, either lithographically defined or "naturally formed" is expected to attenuate tunneling processes responsible for charge leakage, including the dominant PAT mechanism as it requires more photons to transfer charge over larger number of junctions[19]. It is therefore reasonable to assume that the GM film will act as charge leakage suppressor. The details of conduction mechanism in granular films have recently attracted a lot of theoretical attention[44-46]. The common conclusion is that conduction through GM film at low temperatures and small biases is due to tunneling through the virtual states ("multiple cotunneling" [44]) which bypasses simple activation over the Coulomb barriers separating nearest neighbor grains. This mechanism explains why GM films remain conductive even at very low temperatures despite the high values of charging energy of the individual grains. At the same time conductance through GM film at low biases across it is much lower than in any diffusive microstrip resistor, so that any undesired tunneling is strongly suppressed. This is the condition that is required for low leakage rate. On the other hand, once the bias is applied the conductance of the GM dot rises exponentially resulting in fast switching of an electron. Clearly, a more accurate theoretical model is required to describe the details of charge transfer in such a device, particularly to find out the effectiveness of charge leakage suppression and to determine the important parameters of the GM films which can be used to optimize the performance of the devices.

In conclusion, we have fabricated and tested a single-electron QCA latch with a middle dot made of granular metal film ($CrO_x$) connected to the end Al dots with single AlOx tunnel junctions, to investigate the applicability of GM charge leakage suppressors for single-electron logic devices. We observe latching with a leakage error rate, $\Gamma_L \approx 1$ s$^{-1}$, comparable to that obtained for the Al-AlO$_x$ latches with six junctions. The observed charge leakage suppression is provided by the granular media of the central island. The size of the devices with GM leakage suppressors can be made much smaller than by using metallic resistive microstrips, thus alleviating the reduction of charging energy associated with the large self-capacitance of the metallic microstrips. We believe that granular metal films can be used for the fabrication of the devices providing precise charge transfer (pumps and turnstiles) with reduced size and smaller number of lithographically defined tunnel junctions. It also can be potentially used for the fabrication of single-electron memory devices, where strong nonlinearity of the I-V characteristic can be beneficial for short write time and long retention time.

**Acknowledgements.**

The authors wish to thank N. M. Zimmerman for reading the manuscript and numerous valuable suggestions. S. V. Demishev's and N. E. Sluchanko's critical comments are gratefully acknowledged. The work was supported by the MRSEC Center for Nanoscopic Materials Design of the National Science Foundation under Award No. DMR-0080016 and NSF grant CCR-0210153.

Figure Captions:

Figure 1. Sketches of single-electron devices employing granular metal islands. Aluminum (Al) with thin oxide on top is shown in gray, white bars represent CrOx film deposited on top of Al. a) Single electron latch. Middle dot D2 is made of granular metal. Differential input signals ($+V_{IN}$ and $-V_{IN}$), clock signal ($V_{CLK}$), and electrometers (E1 and E2) are capacitively coupled to D1, D2, and D3 as shown in the figure. Electrometers E1 and E2 are fabricated in the second step of e-beam lithography. b) Single-electron transistor with granular metal ($CrO_x$) island. Source and drain leads, and the gate are shown.

Figure 2: Phase plots of single-electron latches: (a, b) Single-electron latch with $CrO_x$ middle dot. Dot potential (of D1) is measured by electrometer E1 for two directions of the differential input signal as shown by the arrows (X axis is the same for both directions of the input voltage). Triangularly shaped areas outlined with dashed lines indicate bistable behavior; (c) An $Al/AlO_x$ latch with the same design (2 junctions). Only one plot is shown because plots for the opposite directions of the scans are indistinguishable in this case. No bistability is seen in this plot.

Figure 3: Operation of the QCA latch with $CrO_x$ middle dot: (a) applied input bias; (b) clock bias; (c) potential of D1 measured by electrometer $E_1$. An instance of a leakage error is shown by the dashed line. (d) Leakage error probability for the latch in the "hold" state as a function of time elapsed from the moment the electron is latched. Cumulative leakage error probability is plotted for 192 scans as a function of time elapsed from the moment the differential input bias is removed ($t_0$, $t_2$). Dashed line is fit using formula (2) with parameter $\tau = 1.05$ s.

Figure 4: Large scale phase plots of the two single-electron latches: (a) Latch with 2 $AlO_x$ junctions and $CrO_x$ middle island, (b) An $Al/AlO_x$ latch with 6 tunnel junctions (two branches connecting the middle dot to the end dots each have 3 junctions). Note that distinct triangular pattern is clearly seen throughout the whole plot for the latch with $CrO_x$ island, whereas it appears only on a small portion of the plot for the $Al/AlO_x$ latch.

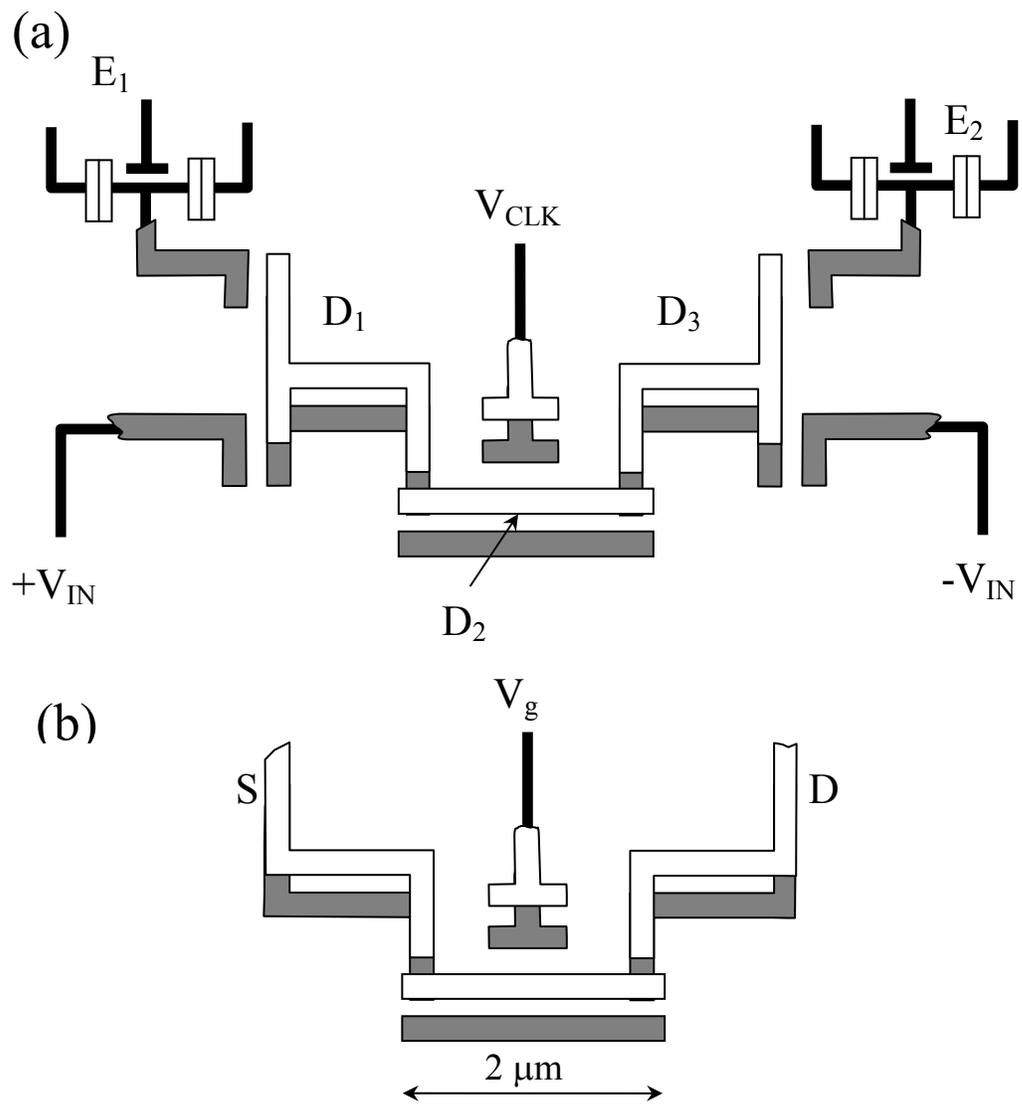

Figure 1.

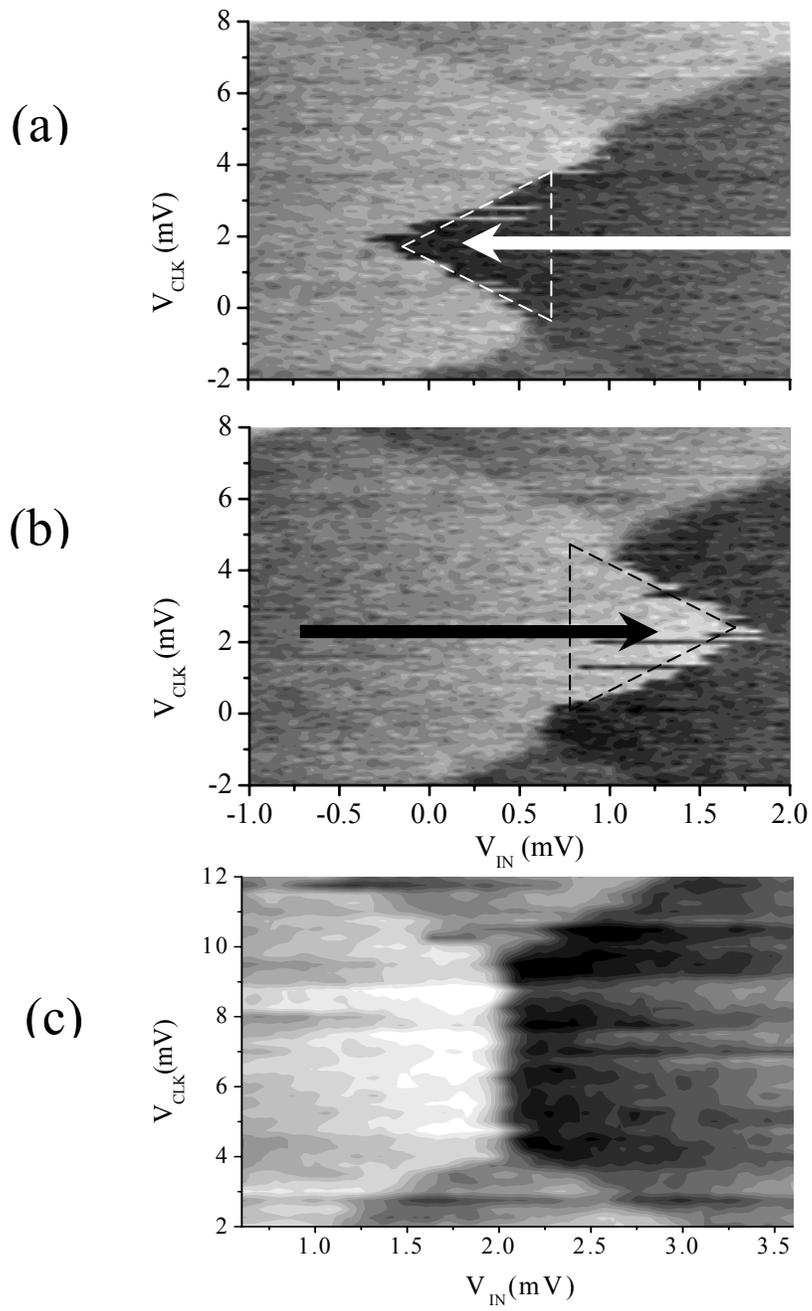

Figure 2.

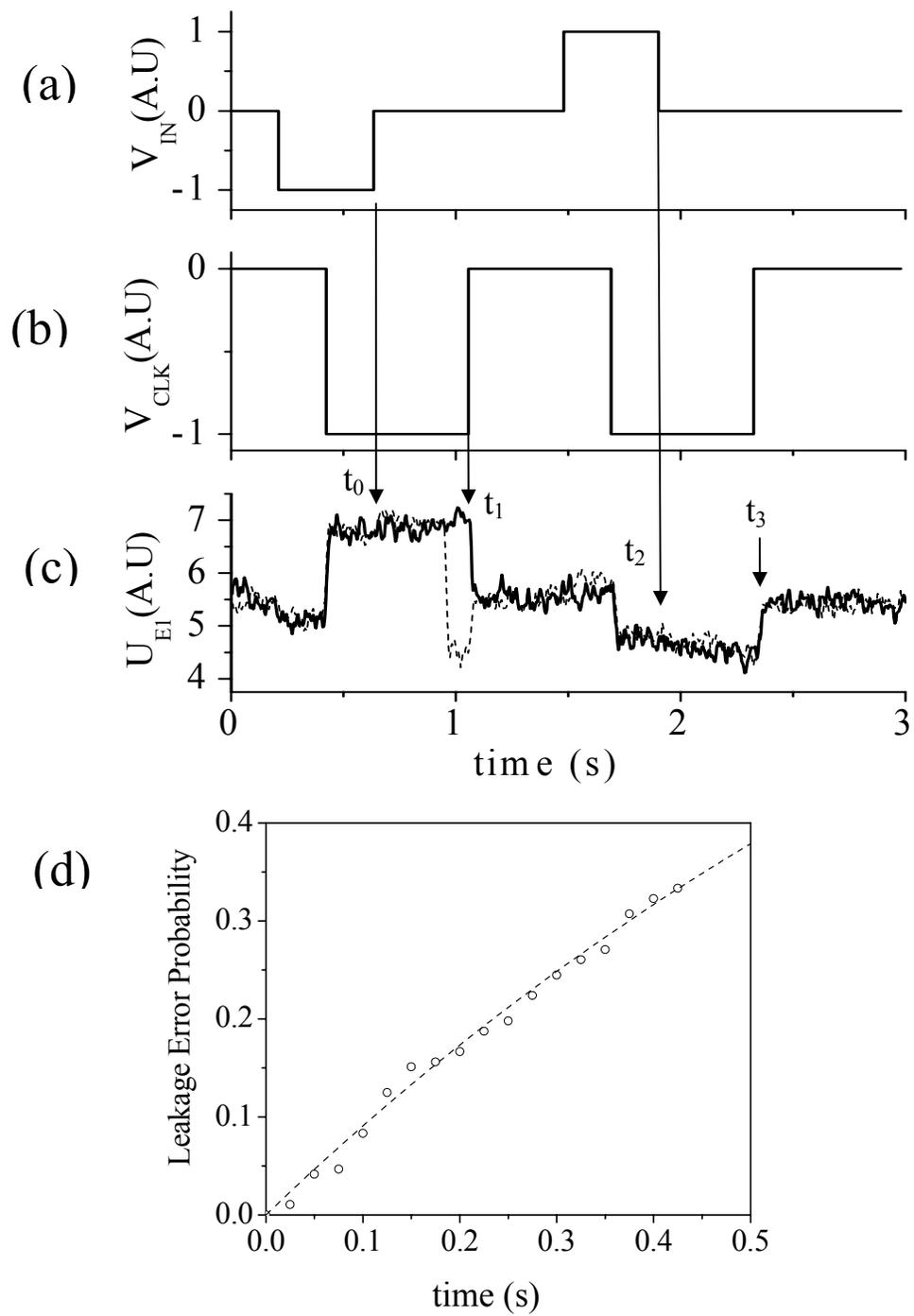

Figure 3.

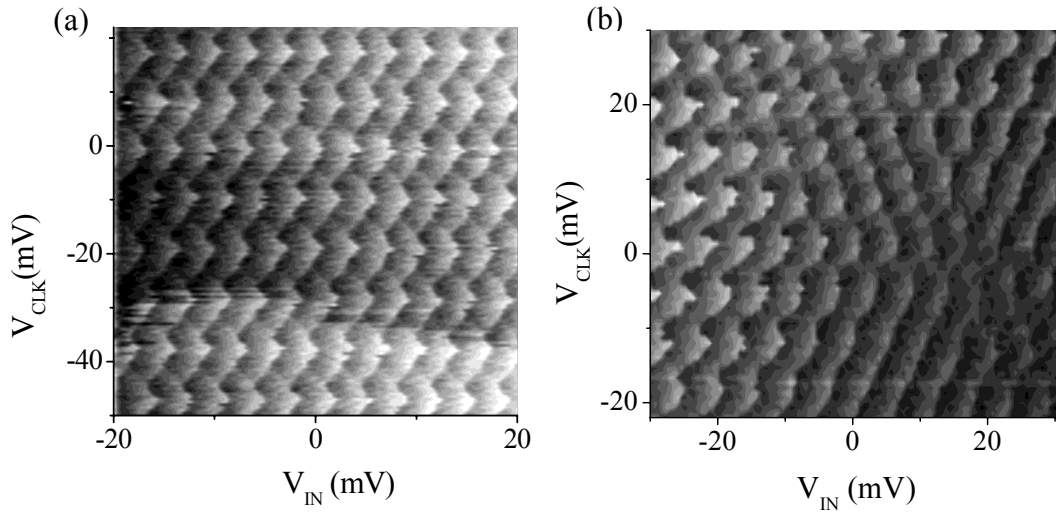

Figure 4.